\documentclass[aps,pre,twocolumn,showpacs]{revtex4}

\usepackage{graphicx}
\usepackage{amsmath}

\begin{document}

\title{Alternative solutions to diluted $p$-spin models and XORSAT
problems}

\author{M. M\'ezard}
\email{mezard@ipno.in2p3.fr}
\affiliation{Laboratoire de Physique Th\'eorique et Mod\`eles
Statistiques, Universit\'e Paris Sud, 91405 Orsay, France}
\author{F. Ricci-Tersenghi}
\email{Federico.Ricci@roma1.infn.it}
\affiliation{Dipartimento di Fisica and INFM, Universit\`a di Roma
``La Sapienza'', Piazzale Aldo Moro 2, I-00185 Roma (Italy)}
\author{R. Zecchina}
\email{zecchina@ictp.trieste.it}
\affiliation{International Center for Theoretical Physics, Strada
Costiera 11, P.O. Box 586, I-34100 Trieste, Italy}

\date{\today}

\begin{abstract}
We derive analytical solutions for $p$-spin models with finite
connectivity at zero temperature.  These models are the statistical
mechanics equivalent of $p$-XORSAT problems in theoretical computer
science.  We give a full characterization of the phase diagram:
location of the phase transitions (static and dynamic), together with
a description of the clustering phenomenon taking place in
configurational space.  We use two alternative methods: the cavity
approach and a rigorous derivation.
\end{abstract}


\maketitle

\section{Introduction}

The very last years have seen a growth of interest in disordered
models defined on Bethe-lattices-like topologies, that is finite
connectivity random graphs (see e.g. Ref.\cite{MEPA}).  Appropriate
generalizations of mean-field theory are exact on such structures
allowing for an exact solution of spin-glass like models.  The
presence of large loops may induce frustration leading to highly non
trivial properties at low enough temperatures.  Interacting models
defined over finite connectivity graphs provide a better approximation
to finite-dimensional models than fully connected mean-field models,
allowing for qualitatively new effects to be discussed.  At zero
temperature, spin glass like models over random graphs correspond to
some random combinatorial optimization problems of central relevance
in theoretical computer science \cite{MAMOZE}.

Quite in general, spin glass models show an interesting phase diagram
in the $(\gamma,T)$ plane (see e.g. Fig.2 in \cite{FMRWZ}), where
$\gamma$ is a parameter proportional to the mean connectivity of the
underlying random graph and $T$ is the temperature.  The frozen phase
is located at high $\gamma$ and/or low $T$.

Open questions are, for example, the exact location of the critical
lines (dynamic and static ones), the full characterization of the
configurational space in the frozen phase (e.g.\ ground state energy
and threshold energies), etc...

Here we focus on the simplest non trivial model that can be defined on
a random graph with finite mean connectivity, namely the $p$-spin
model.  We concentrate on the zero temperature limit, which
corresponds to the $p$-XORSAT problem in theoretical computer science
\cite{XORSAT}.

In this limit the model undergoes two relevant phase transitions
\cite{RIWEZE}.  The first one takes place at $\gamma_d$ (for $p=3$,
$\gamma_d=0.818469$) and corresponds to a clustering phenomenon: For
$\gamma < \gamma_d$ all the ground states (GS) form a unique cluster,
while for $\gamma > \gamma_d$ they split into an exponentially large
(in $N$) number of clusters, each one containing an exponential number
of GS.  This clustering phenomenon coincides with the formation, in
the configurational space, of barriers (clusters are well defined only
because of the presence of barriers) and of metastable states, which
make any greedy search algorithm inefficient.  This is why it is
usually called {\em dynamic transition} \cite{BLRZ}.  The second phase
transition takes place at $\gamma_c > \gamma_d$ (for $p=3$,
$\gamma_c=0.917935$) and marks the SAT/UNSAT transition, that is the
point where frustration becomes manifest in the system and the GS
energy becomes larger than zero.

We will derive the above scenario via two distinct and complementary
methods.  The first one is the {\em cavity method}.  Its power relies
in its generality, since it can be applied easily to more complex
systems too, e.g.\ random k-SAT \cite{MEPAZE,MEZE}.  Within this
method the above scenario can be obtained using an Ansatz with a
single step of replica symmetry breaking (1RSB).  The second method is
a {\em rigorous derivation} based on the `leaf removal' algorithm
which is able to reduce the random (hyper)graph to its relevant {\em
core}.  On the core, any interesting quantity (e.g.\ the number of GS,
cluster size and distance) can be easily calculated, since annealed
averages coincide with quenched ones.

This rigorous derivation is of great importance also because this is
one of the few cases \cite{BOKU} where a highly non trivial scenario,
previously obtained with a replica calculation
\cite{RIWEZE,LERIZE,FLRZ}, can be confirmed with rigorous methods.
These results confirm the validity of the cavity approach, and may
open the way towards the construction of mathematical bases for the
Parisi's replica symmetry breaking theory \cite{MEPAVI}.


\section{Definition of the model}

The random $p$-XORSAT problem consists in finding an assignment to $N$
boolean variables $x_i \in \{0,1\}$, such that a set of $M=\gamma N$
parity checks on these variables are satisfied.  Each parity check is
of the kind
\begin{equation}
x_{i^m_1} + \ldots + x_{i^m_p} = y_m \mod 2, \qquad m = 1, \ldots, M
\end{equation}
where, for each $m$, the $p$ indices $i^m_1, \ldots, i^m_p \in \{1,
\ldots, N\}$ are chosen randomly and uniformly among the
$\binom{N}{p}$ possible $p$-uples of distinct indices, and the
`coupling' $y_m$ takes randomly value 0 or 1 with equal probability.
The above set of constraints can be written in a more compact way as
$\hat{A}\,\vec{x} = \vec{y} \mod 2$, where $\hat{A}$ is a $M \times N$
random sparse matrix with exactly $p$ ones per row and $y$ is a random
vector of $0$s and $1$s.

Once the mapping $\sigma=(-1)^x$ and $J=(-1)^y$ is performed, the
XORSAT problem can be also studied as the zero-temperature limit of
the following $p$-spin Hamiltonian giving the energy for a
configuration of $N$ Ising spins $\sigma_i \in \{-1,1\}$:
\begin{equation}
\mathcal{H} = \sum_{m=1}^M \left( 1 - J_m \sigma_{i_1^m} \ldots
\sigma_{i_p^m} \right) \quad .
\label{eq:ham}
\end{equation} 
Unfrustrated ground states (GS) configurations have zero energy and
correspond to solutions of the XORSAT problem, since they satisfy all
the constraints: $\forall m \; , \; \sigma_{i_1^m} \ldots
\sigma_{i_p^m} = J_m $.


\section{$T=0$ phase diagram from the one-step cavity method}
\label{sec:cavity}

In this section we shall display the analysis of the phase diagram of
the $p$-spin problem as it arises from the one-step cavity approach.
We consider the cavity formalism directly at zero temperature as
discussed by M\'ezard and Parisi \cite{MEPA,MEPA_T=0} and developed
further in \cite{MEZE}. We refer to those papers for a review of the
method and the notations. Here we shall limit ourselves to the
technical aspects of the analytical calculation for $p=3$ case,
generalizations to $p>3$ being straightforward.

The zero temperature $p$-spin model can be viewed as a relatively
simple limit case of more general problems such as random k-SAT for
which the cavity calculations have also been carried out recently
\cite{MEZE}. The main technical difference between random k-SAT like
problems and the $p$-spin model consists in the fact that the site
dependence of the functional order parameter simplifies dramatically
in the $p$-spin problem below the static transition. This allows for a
rigorous derivation of the cavity and replica results by alternative
methods, as we shall thoroughly discuss in the subsequent sections.

\begin{figure}[!ht]
\includegraphics[width=0.8\columnwidth]{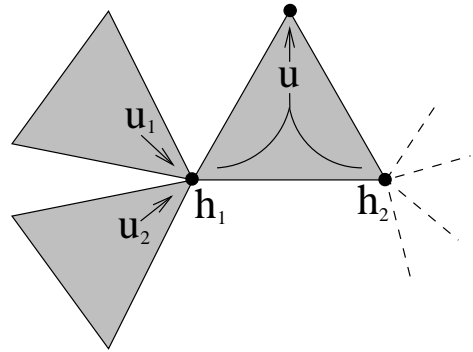}
\caption{A pictorial view of the cavity iteration: $h_1$ and $h_2$
cavity fields are the sum of some cavity biases $u$, and in turn
they generate a new cavity bias $u$ according to Eq.~(\ref{uwdef2}).}
\label{propa}
\end{figure}

In the cavity formalism \cite{MEPAVI} one works with ``cavity fields''
$h_i$ associated to the sites and ``cavity biases'' $u_J$ associated
to the hyperedges.  The cavity field is the effective field on a
variable once one of its interactions has been removed.  Under a
cavity iteration, cavity biases generate cavity fields and vice versa
(see Fig.~\ref{propa}).  The cavity field $h$ is always the sum of the
cavity biases $u$ coming from all its interactions, but the one
removed.  The rule for generating $u$ biases from $h$ fields is in
general more complex.

For $T=0$ the formalism simplifies a lot \cite{MEPA_T=0}: Cavity
fields and cavity biases only take integer values and the cavity
equations can be derived easily by implementing the energy
minimization condition under the cavity iteration.  Let us imagine to
add a hyperedge connecting 3 spins, say spins $\sigma_0$, $\sigma_1$
and $\sigma_2$ among which spin $\sigma_0$ plays the role of cavity
spin.  We need to perform a partial minimization of the effective
energy
\begin{multline}
\min_{\sigma_1,\sigma_2} \left[ \epsilon( \sigma_0, \sigma_1,
\sigma_2) - \left( h_1 \sigma_1 + h_2 \sigma_2 \right) \right] = \\
= - w_J(h_1,h_2) - u_J(h_1,h_2)\; \sigma_0 \quad ,
\label{uwdef2}
\end{multline}
where $\epsilon( \sigma_0, \sigma_1, \sigma_2 ) = 1 - J \sigma_0
\sigma_1 \sigma_2$.  The above relation defines the cavity biases $w$
and $u$ as functions of the ``input'' cavity fields $h$.  After a
little algebra one finds
\begin{eqnarray}
w_J(h_1,h_2) &=& |h_1| + |h_2| - |u_J(h_1,h_2)| \ , \nonumber \\
u_J(h_1,h_2) &=& {\cal S}(J h_1 h_2) \ ,
\label{wu_3spin}
\end{eqnarray}
where the function ${\cal S}(x)$ is defined as
\begin{equation}
{\cal S}(x) = \left\{
\begin{array}{cc}
\text{sign}(x) & \text{ if\ \ } x \ne 0 \quad ,\\
0 & \text{ if\ \ } x = 0 \quad .
\end{array}
\right.
\end{equation}
The free-energy of the system can be expressed either in terms of
probability distributions of the cavity fields or of the cavity biases
\cite{MEPA,MEPA_T=0,MEZE}.

In a one step scenario the phase space breaks into many pure states
and the order parameter of the model is a complete histogram, over the
system, of probability distribution functions of fields, ${\cal
P}[P(h)]$, and biases, ${\cal Q}[Q(u)]$.  Such a rich structure of the
order parameter can be understood by noticing that each spin may
fluctuate from state to state and therefore the whole collection of
single site probability distributions might be needed to capture such
fluctuations.  In the simple case of a single pure state, the so
called replica symmetric (RS) phase, single site probability
distributions becomes delta functions and the order parameter
simplifies to a single global probability distribution.

Following the general scheme discussed in
Refs.~\cite{MEPA,MEPA_T=0,MEZE,MEPAZE}, but with a more convenient
normalization for the $Q(u)$, the self-consistency equation for the
${\cal Q}[Q(u)]$ reads
\begin{widetext}
\begin{eqnarray}
Q(u) &=& E_J \int dh P^{(k)}(h)\; dg P^{(k')}(g)\; \delta\Big( u -
u_J(h,g) \Big) \qquad \text{with prob.} \quad e^{-3\gamma}
\frac{(3\gamma)^k}{k!}\; e^{-3\gamma} \frac{(3\gamma)^{k'}}{k'!} \ ,
\nonumber \\
P^{(k)}(h) &=& \frac{1}{A_k} \int du_1 Q_1(u_1) \ldots du_k Q_k(u_k)
\; \delta \left( h - \sum_{i=1}^k u_i \right) \exp\left[-y
\left(\sum_{i=1}^k |u_i| - |\sum_{i=1}^k u_i|\right)\right] \ ,
\label{QRSBequation}\\
A_k &=& \int du_1 Q_1(u_1) \ldots du_k Q_k(u_k) \exp\left[-y
\left(\sum_{i=1}^k |u_i| - |\sum_{i=1}^k u_i|\right)\right] \ ,
\nonumber
\end{eqnarray}
\end{widetext}
where all the $Q_i(u)$ on the r.h.s. are chosen randomly from the
distribution ${\cal Q}[Q(u)]$.  The average $E_J$ over the coupling
signs $J=\pm1$ forces all the distribution to be symmetric under $u
\leftrightarrow -u$ or $h \leftrightarrow -h$.  The parameter $y$ is
the so called reweighting coefficient ($y=\beta m$ where $m$ is the
Parisi breaking parameter) which takes into account level crossing of
states under the cavity iterations \cite{MEPA_T=0}.  The parameter $y$
must be chosen such as to maximize the free energy.

As the cavity biases take values in $\{0,\pm 1\}$, and thanks to above
mentioned symmetry, each $Q_i(u)$ can be written, in full generality,
as
\begin{equation}
Q_i(u) = \eta_i\, \delta(u) + \frac{1-\eta_i}{2} \big[ \delta(u+1) +
\delta(u-1) \big] \ .
\end{equation}
Thus the self-consistency equation for ${\cal Q}[Q(u)]$ can be
rewritten as a self-consistency equation for the probability
distribution of $\eta_i$, $\rho(\eta)$.

Eventually, given the whole set of stationary $\{ Q_i(u) \}$ or the
stationary $\rho(\eta)$, the average ground state energy and the
complexity can be deduced from the formulae of
Refs.~\cite{MEPA,MEPA_T=0,MEZE,MEPAZE}.

\subsection{Solution of the self-consistency equation}

\subsubsection{RS solution}

We first notice that it is always possible to get back the simple
replica symmetric solution by fixing $y=0$ and assuming that the
cavity biases are ``certain'', $Q_j(u)=\delta(u-u_j)$, where the $u_j$
are independent and identically distributed random variables taken
from a distribution
\begin{equation}
\mathcal{Q}(u) = c_0\;\delta(u) + \frac{(1-c_0)}{2} \big[ \delta(u-1)
+ \delta(u+1) \big] \quad .
\label{Qc0}
\end{equation}
Plugging the above form into Eqs.~(\ref{QRSBequation}), one finds for
$c_0$
\begin{multline}
1-\sqrt{1-c_0} = \\
= e^{-3\gamma} \sum_k \frac{(3\gamma)^k}{k!}  \sum_{q=0}^{\lfloor k/2
\rfloor} \binom{k}{2q} \binom{2q}{q} \left(\frac{1-c_0}{2}\right)^{2q}
c_0^{k-2q} = \\
= e^{-3\gamma(1-c_0)} I_0\left[3\gamma(1-c_0)\right] \quad ,
\label{p0}
\end{multline}
where $\lfloor x \rfloor$ is the integer part of $x$.  However, the
above equation leads to wrong predictions: a solution different from
the trivial paramagnetic one, $Q_j(u)=\delta(u)$, appears at
$\gamma=1.16682$ with a negative energy.  At $\gamma_{RS}=1.29531$ the
energy becomes positive, giving a lower bound for the true energy of
the system.

\subsubsection{1RSB solution and the existence of non trivial fields}

The numerical solution of Eq.~(\ref{QRSBequation}) indicates that
there exits a non-trivial solution in the region $\gamma \agt 0.82$
for sufficiently large values of the reweighting $y$.  A careful look
at the numerics shows that the probability distributions of $\eta_i$
takes the form
\begin{equation}
\rho(\eta) = t\: \delta(\eta-1) + (1-t)\: \tilde\rho(\eta) \quad ,
\label{Qshape3spin}
\end{equation}
that is a fraction $t$ of cavity biases are trivial.  The non trivial
cavity biases are characterized by a distribution $\tilde\rho$ which
shrinks in the limit of large $y$, converging to delta function in
$\eta=0$.  The $y \to \infty$ limit is particularly relevant in the
region up to $\gamma_c$.

\subsubsection{The $y\to\infty$ limit: the complexity and the location
of the phase transition}

Looking at the self-consistency equations (\ref{QRSBequation}), the
only way one can obtain a non-trivial distribution $Q(u)$ on the
l.h.s. is when both $P^{(k)}(h) \neq \delta(h)$ and $P^{(k')}(g) \neq
\delta(g)$.  Moreover the probability that $P^{(k)}(h) = \delta(h)$
equals the probability of picking up $k$ trivial distributions $Q(u)$,
i.e. $t^k$.  Putting everything in formulae, one has
\begin{multline}
1-t = e^{-6\gamma} \sum_{k,k'=0}^\infty \frac{(3\gamma)^k}{k!}
\frac{(3\gamma)^{k'}}{k'!} (1-t^k) (1-t^{k'}) \ \Longrightarrow \\
1-t = \left(1 - e^{-3 \gamma (1-t)} \right)^2 \ . \qquad
\label{EQt3spin}
\end{multline}
For $\gamma < \gamma_d =0.818469$ the only solution is $t=1$ (the system
is a paramagnet) whereas above $\gamma_d$ a non-trivial solution
appears.

For $y=\infty$, a direct inspection of the numerical results shows
that the cavity biases spontaneously divide in two categories, such
that $\rho(\eta) = t\, \delta(\eta-1) + (1-t)\, \delta(\eta)$.
In terms of ${\cal Q}[Q(u)]$ it corresponds to having
\begin{equation}
Q(u)= \left\{
\begin{array}{ccc}
\delta(u) & \text{ with prob. } & t\\
\frac{\delta(u-1) + \delta(u+1)}{2} & \text{ with prob. } & 1-t
\end{array}
\right.
\label{Ansatz_3spin_1}
\end{equation}
which indeed is a fixed point under the iteration process
(\ref{QRSBequation}) for $y =\infty$, provided the fraction of trivial
biases $t$ satisfies Eq.~(\ref{EQt3spin}).

\begin{figure}[!ht]
\includegraphics[width=\columnwidth]{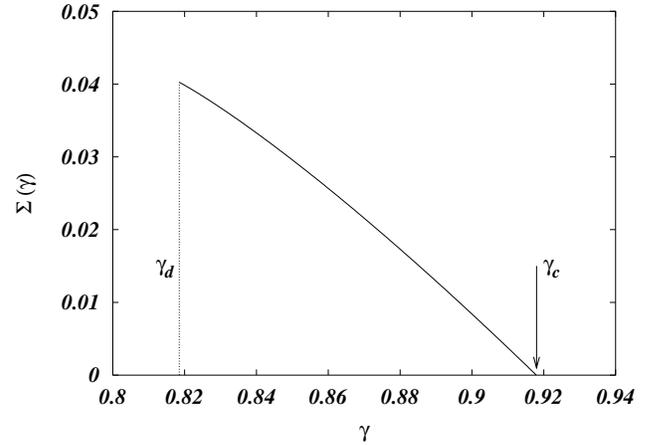}
\caption{The complexity as a function of $\gamma$.}
\label{sigma}
\end{figure}

Using the expressions of Refs.~\cite{MEPA,MEPA_T=0,MEZE,MEPAZE}, for
very large $y$, the free energy can be written as $\Phi(y) =
\frac{\psi}{y}$.  As expected, one finds that, as long as $\psi<0$ the
maximum of $\Phi(y)$ is located in $y=\infty$ and corresponds to a
zero ground state energy.  Consequently the complexity or
configurational entropy of zero-energy states, i.e. the normalised
logarithm of the number of solutions clusters is given by
\begin{equation}
\Sigma = -\psi = \log(2) \left[ 1 -\frac\lambda3 -e^{-\lambda} \left(1
+\frac23 \lambda\right)\right] \quad ,
\end{equation}
where $\lambda=3\gamma(1-t)$ and satisfies the self-consistency
equation
\begin{equation}
\lambda = 3\gamma \left(1-e^{-\lambda}\right)^2 \quad .
\label{eqlam}
\end{equation}
The critical point, i.e.\ the SAT/UNSAT threshold, $\gamma_c=0.917935$
can be found as the $\gamma$ value where the complexity becomes zero
(see Fig.~\ref{sigma}).  For $\gamma>\gamma_c$ the free energy
$\Phi(y)$ has a positive maximum in a finite value of $y$, which
corresponds to a positive ground state energy.

\subsubsection{Expansion at large $y$: the ground state energy in the
UNSAT phase (MAX-3-XORSAT)}

In order to study the ground state energy for $\gamma>\gamma_c$ we
need to take care of the leading corrections in the limit $y \gg 1$.
For finite $y$, the distribution in Eq.~(\ref{Ansatz_3spin_1}) is no
longer stable and we need to study a more general distribution of
biases which takes care of the appearance of a non-trivial
contribution to the peak in $u=0$ arising from frustrated
interactions.  This more general ${\cal Q}[Q(u)]$ is such that a
fraction $t$ of messages is still completely trivial,
$Q(u)=\delta(u)$, while non-trivial messages comes from distributions
of the following kind
\begin{equation}
Q(u) = \xi e^{-2y}\, \delta(u) + \frac{1-\xi e^{-2y}}{2} \big[
\delta(u-1) + \delta(u+1) \big] \ .
\label{Ansatz_3spin_2}
\end{equation}
The factor $e^{-2y}$ has been introduced in order to have finite $\xi$
in the limit of very large $y$.  Moreover, from numerical solution of
Eq.~(\ref{QRSBequation}) we observe that $\xi$ takes only integers
values.  Let us call $a_m$ the fraction of non-trivial distributions
having $\xi=m$.  The generating function $a(z)=\sum_m a_m z^m$
satisfies the equation
\begin{equation}
a(z) = \left[ A\, a(z) + B\, z +1-A-B \right]^2 \ ,
\label{eqa}
\end{equation}
where $A=\frac{e^{-\lambda} \lambda}{1-e^{-\lambda}}$ and
$B=\frac{e^{-\lambda} \lambda^2/2}{1-e^{-\lambda}}$.
 
Using the distributions in Eq.~(\ref{Ansatz_3spin_2}) one can obtain
the free energy density $\Phi(y)$ up to the first correction
\begin{equation}
\Phi(y) = \frac{\psi}{y} -\frac{\omega}{y} e^{-2y} +
\mathcal{O}(e^{-4y}) \ ,
\label{phi3spinrsb}
\end{equation}
where
\begin{multline}
\omega = \frac{\lambda}{3} \Bigg[1-e^{-\lambda} \Big(1+\frac32\lambda
+\lambda^2\Big) + \\
+ \langle\xi\rangle \Big(1-e^{-\lambda}(1+2\lambda) \Big) \Bigg]
\quad .
\end{multline}
The mean value of $\xi$ can be easily obtained deriving
Eq.~(\ref{eqa}) with respect to $z$ and then putting $z=1$,
\begin{equation}
\langle\xi\rangle = a'(1) = \frac{2B}{1-2A} =
\frac{e^{-\lambda} \lambda^2}{1-e^{-\lambda} (1+2\lambda)} \ ,
\end{equation}
and thus we have
\begin{equation}
\omega = \frac{\lambda}{3} \Bigg(1 -e^{-\lambda} \bigg(1 +\frac32
\lambda \bigg) \Bigg) \quad .
\end{equation}

Summarizing the statistical mechanics analysis, we have that for any
$\gamma>\gamma_d=0.818469$, one can solve Eq.~(\ref{eqlam}) for
$\lambda$, deduce the large $y$ behaviour of $\Phi(y)$ from
Eq.~(\ref{phi3spinrsb}) and maximize $\Phi(y)$ with respect to $y$.
We find a critical value of $\gamma$, $\gamma_c=0.917935$, where
$\psi$ changes sign.  For $\gamma<\gamma_c$, $\psi<0$ and therefore
the maximum of $\Phi(y)$ is found at $y=\infty$. The distribution of
cavity biases is given by Eq.~(\ref{Ansatz_3spin_1}), and the maximum
value of $\Phi$ is $0$, showing that all hyperedges are satisfied
(apart from maybe a vanishing fraction at large $N$).

At $\gamma =\gamma_c$, $\psi$ changes sign and, for $\gamma >
\gamma_c$, $\Phi(y)$ has a maximum at a finite $y$, which shows that
the ground state energy becomes strictly positive: It is no longer
possible to satisfy simultaneously all the constraints.

The value of the energy for $\gamma$ slightly above $\gamma_c$ can be
computed from the large $y$ expansion. Moreover, such an expansion
allows us to compute the complexity $\Sigma(E)$ of states of given
energy $E$ by a Legendre transformation of the free energy.  The
complexity function $\Sigma(E)$ is obtained by solving $E=\partial_y
(y \Phi)$ and $\Sigma = y^2 \partial_y \Phi$.

From Eq.~(\ref{phi3spinrsb}) for $\Phi$ we get
\begin{eqnarray}
E &=& 2 \omega e^{-2y} + \mathcal{O}(e^{-4y}) \quad , \\
\Sigma &=& - \psi + (2 y + 1) \omega e^{-2y} + \mathcal{O}(e^{-4y})
\quad .
\end{eqnarray}
For $\gamma_d<\gamma<\gamma_c$, the constant $\psi$ is negative and
one finds a complexity curve which starts positive at $E=0$
\begin{eqnarray}
\Sigma(E) \simeq - \psi - \frac{E}{2} \left[ \log \left(
\frac{E}{2\omega} \right) - 1 \right] \quad .
\label{sigmares3spin}
\end{eqnarray}
In particular, the number of lowest lying states, which have an energy
$E=0$, scales with the number of $N$ of spins as $\exp(-N \psi)$.

\begin{figure}[!ht]
\includegraphics[width=\columnwidth]{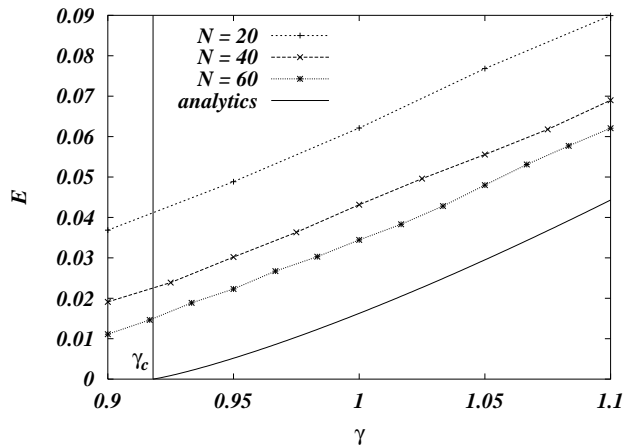}
\caption{Ground state energy for $\gamma$ values above the critical
point $\gamma_c=0.917935$. Numerical data seem to converge to the
analytic prediction. Finite-size corrections roughly decrease as
$1/N$.}
\label{num_ana}
\end{figure}

For $\gamma> \gamma_c$, the expression (\ref{sigmares3spin}) for the
complexity still holds, but $\psi$ is positive. The regime of energies
close to $0$ where $\Sigma(E)$ is negative corresponds to a region
where the average number of states is exponentially small in $N$.
Therefore there are no states in this region in the typical sample.
States appear above the ground state energy $E_0$ which is the point
where $\Sigma(E)$ vanishes, and corresponds to the maximum of
$\Phi(y)$.  In Fig.~\ref{num_ana} we show the analytic prediction for
the ground state energy $E_0$ (lowest curve) together with numerical
results from exact optimization on small systems.  Numerical data are
compatible with the analytic solution, which has been obtained
expanding around the critical point.

\section{Rigorous derivation of thresholds and clustering}
\label{sec:rig}

We now show how the results of the previous Section can be rederived
in a rigorous way.  We will exploit concepts from graph theory and all
the calculations will be simple annealed averages, which are rigorous.
All the formulas will be written for generic $p$, and the particular
case $p=3$ will be considered in order to make connection with
calculations in the Section~\ref{sec:cavity}.

The physical idea behind the graph theoretical derivation is the
following.  In a random hypergraphs there are many variables with
connectivities 0 and 1, whose cavity fields are null.  A small
fluctuation in the number of these variables, induce very large
fluctuations in physical observables, like e.g.\ in the entropy.  Thus
the idea is to remove all these spins and to study the properties of
the residual hypergraph, the core.  We find that, on the core,
sample-to-sample fluctuations are negligible and this allow us to
study its properties by mean of very simple annealed averages.

The plan of this section is the following: (A) definition of some
graph theoretical concepts, like random hypergraph and hyperloop; (B)
introduction of the `leaf removal' algorithm and solution to its
dynamics (estimation of the $\gamma_d$ threshold); (C) statistical
description of the hypergraph core (the part left by the application
of leaf removal algorithm); (D) calculation of $\gamma_c$, the
SAT/UNSAT threshold; (E) derivation of GS clustering properties.

\subsection{Random hypergraphs and hyperloops}

In the Hamiltonian (\ref{eq:ham}) disorder enters in 2 ways: in the
sign of the couplings $J_m=\pm1$ and in the $M$ random $p$-uples of
indices $\{i_1^m,\ldots,i_p^m\}_{m=1,\ldots,M}$, which define the
interactions topology.  This topology has finite connectivity (each
variable appears on average in $p \gamma$ interactions) and locally
tree-like (an Husimi tree for $p>2$).

This topology can be represented as a {\em hypergraph} $\mathcal{G}$
made of a set of $N$ vertices (corresponding to the variables in the
problem) and a set of $M$ hyperedges (corresponding to the constraints
in the problem), each one connecting $p$ vertices.  The disorder
ensemble thus corresponds to all the possible ways one can place
$M=\gamma N$ hyperedges among $N$ vertices, each hyperedge connecting
$p$ vertices and carrying a random sign $J_m=\pm1$.

Analogously to what happens with loops in usual graphs ($p=2$), in a
disordered model defined on a hypergraph ($p>2$) frustration is
induced by the presence of hyperloops~\cite{RIWEZE,LERIZE}, which are
also called hypercycles in the literature~\cite{KOLCHIN}.  The
definition of a hyperloop can be given both in terms of the hypergraph
$\mathcal{G}$ or in terms of the matrix $\hat{A}$.

A hyperloop is a sub-hypergraph $\mathcal{C} \subset \mathcal{G}$,
i.e.\ a set of hyperedges belonging to $\mathcal{G}$, such that every
vertex has even degree (connectivity) in $\mathcal{C}$.

In terms of the matrix $\hat{A}$ it corresponds to a set of rows
$\mathcal{R}$ such that, for every column, the sum modulo 2 of the
elements is zero, i.e.\ $\sum_{i \in \mathcal{R}} A_{ij} \mod 2 = 0 \
\forall j$.

The presence of hyperloops is directly related to the presence of
frustration in the system: If the product of the signs of hyperloop
interactions is negative, $\prod_{m \in \mathcal{C}} J_m = -1$, then
not all such interactions can be satisfied at the same time.  The
critical point $\gamma_c$, where hyperloops percolate, is a $T=0$
phase boundary for the $p$-spin glass models defined by Hamiltonian
(\ref{eq:ham}): For $\gamma < \gamma_c$ all the interactions can be
satisfied and the GS energy is zero, while for $\gamma > \gamma_c$ the
system is in a frustrated spin glass phase and GS of zero energy no
longer exist.

The critical point $\gamma_c$ corresponds to the SAT/UNSAT threshold
for the random $p$-XORSAT problem.  In terms of the random linear
system $\hat{A}\,\vec{x} = \vec{y} \mod 2$, as long as $\gamma <
\gamma_c$, solutions to the system will exist with probability $1$ in
the large $N$ limit {\em for any} $y$.

\subsection{`Leaf removal' algorithm}

\begin{figure}[!ht]
\includegraphics[width=0.8\columnwidth]{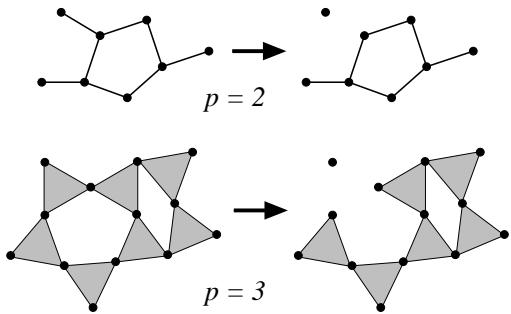}
\caption{A single step of the `leaf removal' algorithm on a graph
(top) and on a hypergraph (bottom).}
\label{one_step}
\end{figure}

Given a hypergraph the leaf removal algorithm proceeds as follows
\cite{PISPWO}: As long as there is a vertex of degree 1 remove its
unique hyperedge.  A single step of the algorithm is illustrated in
Fig.~\ref{one_step} for a graph ($p=2$) and for a hypergraph ($p=3$).
Very similar algorithms have been recently studied in
\cite{BAGO,WEIGT}.

During the whole process the remaining hypergraph is still a random
one, since no correlation can arise among the hyperedges if it was not
present at the beginning.  When there are no more vertices of degree 1
in the hypergraph the process stops and we call {\em core} the
resulting hypergraph, cleared of all isolated vertices.

The leaf removal algorithm is not able to break up any hyperloop,
since each vertex in the hyperloop has at least degree 2.  The
$\gamma$ value where the core size becomes different from zero, let us
call it $\gamma_d$, is certainly smaller than the percolation point of
hyperloops $\gamma_c$ (for $p=2$ these two values coincide).

The evolution of a hypergraph under the application of the leaf
removal algorithm can be described in terms of the probability,
$f_k(t)$, of finding a vertex of degree $k$ after having removed $tN$
hyperedges where the `time' $t$ ranges from 0 to $\gamma$.  The
initial condition is $f_k(0) = e^{-p\gamma} \frac{(p\gamma)^k}{k!}$
and the evolution equations read (see Ref.~\cite{WEIGT} for a detailed
derivation of similar equations)
\begin{eqnarray}
\frac{\partial f_0(t)}{\partial t} &=& (p-1)\frac{f_1(t)}{m(t)} + 1
\quad , \nonumber\\
\frac{\partial f_1(t)}{\partial t} &=& (p-1)\frac{2 f_2(t) - f_1(t)}
{m(t)} - 1 \quad , \label{eq:evol}\\
\frac{\partial f_k(t)}{\partial t} &=& (p-1) \frac{(k+1) f_{k+1}(t) -
k f_k(t)}{m(t)} \qquad \forall k \ge 2 \quad , \nonumber
\end{eqnarray}
where $m(t)=\sum_k k f_k(t) = p (\gamma - t)$, since the mean degree
linearly decreases with time (we remove one interaction per step) and
vanishes at $t=\gamma$.

Thanks to the simplicity of the leaf removal process, the degree
distribution always remains Poissonian for degrees larger than 1, with
a time dependent average $\lambda(t)$,
\begin{equation}
f_k(t) = e^{-\lambda(t)} \frac{\lambda(t)^k}{k!} \qquad \forall k \ge
2 \quad .
\end{equation}
The solution to Eqs.~(\ref{eq:evol}) reads
\begin{eqnarray}
\lambda(t) &=& p \left[ \gamma (\gamma-t)^{p-1} \right]^\frac1p \quad
, \label{eq:lambda}\\
f_1(t) &=& \lambda(t) \left[ e^{-\lambda(t)} - 1 + \left(
\frac{\lambda(t)}{p\gamma} \right)^\frac1{p-1} \right] \quad ,
\label{eq:f1}\\
f_0(t) &=& 1 - \sum_{k=1}^\infty f_k(t) \quad .
\end{eqnarray}
The leaf removal algorithm stops when there are no more vertices of
degree 1, so one can predict the resulting core by fixing $\lambda(t)
= \lambda^*$, where $\lambda^*$ is the largest zero of the equation
$f_1=0$ or equivalently
\begin{equation}
e^{-\lambda^*} - 1 + \left( \frac{\lambda^*}{p\gamma}
\right)^\frac1{p-1} = 0 \quad .
\label{eq:ls}
\end{equation}
More precisely $\lambda^*$ is the first zero of Eq.~(\ref{eq:f1}) one
finds decreasing $\lambda$, starting from the initial value of
$\lambda(0)=p\gamma$, but this always coincides with the largest zero.
Note that once we define $m=[\lambda^*/(p\gamma)]^{1/(p-1)}$,
Eq.~(\ref{eq:ls}) can be rewritten as
\begin{equation}
1 - m = \exp\left(-p \gamma m^{p-1} \right) \quad ,
\label{eq:mag}
\end{equation}
which is nothing but the equation for the magnetization in the
ferromagnetic state \cite{RIWEZE}, equivalently the equation for the
backbone size in any cluster.  Note that Eq.~(\ref{eq:ls}) with $p=3$
is identical to Eq.~(\ref{eqlam}), which indeed determines the mean
connectivity of the sub-hypergraph made of hyperedges with non-trivial
biases.

\begin{figure}[!ht]
\includegraphics[width=\columnwidth]{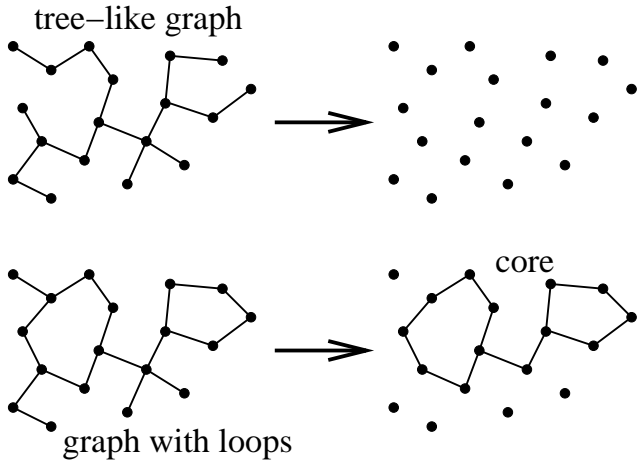}
\caption{On graphs ($p=2$) the leaf removal algorithm is not able to
break loops, which thus remain in the residual core.}
\label{lr2}
\end{figure}

\begin{figure}[!ht]
\includegraphics[width=\columnwidth]{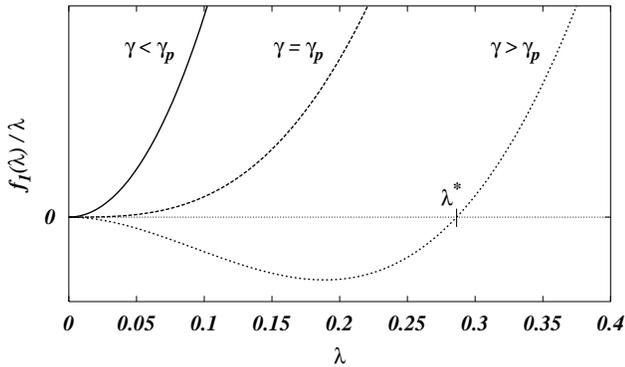}
\caption{The function $f_1(\lambda)/\lambda$ for $p=2$.}
\label{eq2}
\end{figure}

In the $p=2$ case the leaf removal algorithm is able to delete all the
edges only for tree-like graphs.  As soon as there are loops in the
graph, a core containing these loops arises (see Fig.~\ref{lr2}).  In
a random graph the leaf removal transition coincides with the
percolation one at $\gamma_p=1/2$.  The shape of the function
$f_1(\lambda)$ is shown in Fig.~\ref{eq2}: For $\gamma \le \gamma_p$,
there is only one zero in $\lambda^*=0$; While, for $\gamma >
\gamma_p$, $\lambda^* > 0$ and a core arises, whose size grows as
$(\gamma - \gamma_p)^2$ near the critical point.

\begin{figure}[!ht]
\includegraphics[width=\columnwidth]{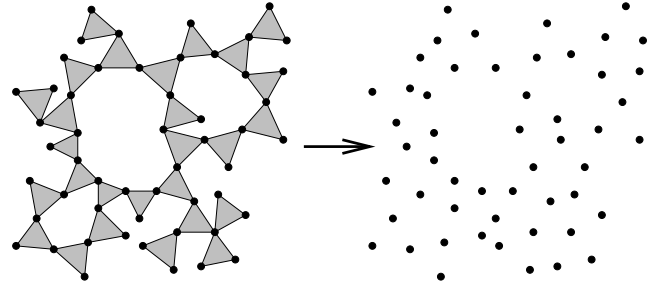}
\caption{For $p>2$ the leaf removal algorithm is able to break loops
(but not hyperloops!).}
\label{lr3}
\end{figure}

For $p>2$ the percolation transition, taking place at
$\gamma_p=\frac{1}{p(p-1)}$, does not affect at all the leaf removal
algorithm which is able to delete all the hyperedges, even those
forming loops (but not those forming hyperloops!), far beyond
$\gamma_p$ (see Fig.~\ref{lr3}).

\begin{figure}[!ht]
\includegraphics[width=\columnwidth]{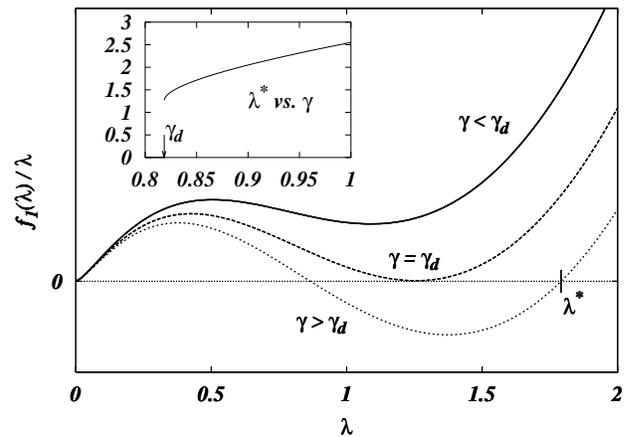}
\caption{The function $f_1(\lambda)/\lambda$ for $p=3$. Inset:
function $\lambda^*(\gamma)$ for $p=3$.}
\label{eq3}
\end{figure}

The leaf removal transition takes place at $\gamma_d$, which is
defined as the first $\gamma$ value where a second solution to
Eq.~(\ref{eq:mag}) appears.  For $p=3$ we have $\gamma_d=0.818469$.
The transition is first order and, at the critical point, the core
already occupies a finite fraction of the system.  In Fig.~\ref{eq3}
we show the function $f_1(\lambda)$ for $p=3$.  It is clear (see inset
of Fig.~\ref{eq3}) that when $\lambda^*(\gamma)$ becomes different
from zero it directly jumps to a finite value:
$\lambda^*(\gamma_d)=1.25643$ for $p=3$.

\subsection{Statistical description of the core}

\begin{figure}[!ht]
\includegraphics[width=\columnwidth]{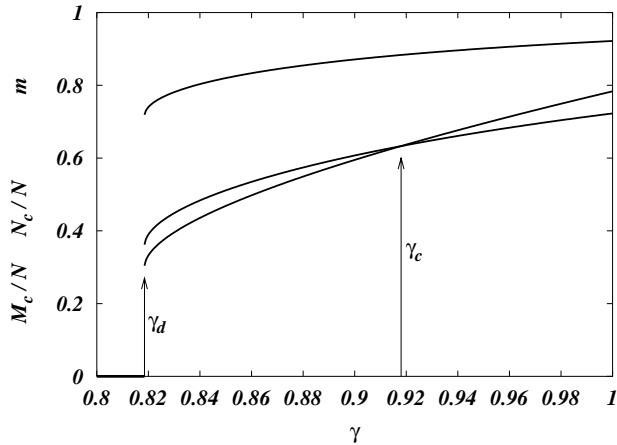}
\caption{From bottom to top (on the left): For $p=3$, normalized
number of hyperedges and vertices in the core, and fraction of frozen
sites, i.e. magnetization (or backbone) in a state.}
\label{p3}
\end{figure}

Once the leaf removal process has come to an end the distribution of
connectivities on the core is a truncated Poissonian
\begin{equation}
P_c(k) = \left\{
\begin{array}{ll}
0 & \text{for } k=0,1\\
e^{-\lambda^*(\gamma)} \frac{\lambda^*(\gamma)^k}{k!} & \text{for } k
\ge 2
\end{array}
\right .
\label{eq:Pk}
\end{equation}
The number of vertices $N_c$ and the number of hyperedges $M_c$ in the
core can be expressed in terms of $p$, $\gamma$ and
$\lambda^*(\gamma)$ as
\begin{widetext}
\begin{eqnarray}
N_c(\gamma) &=& N \sum_{k=2}^\infty f_k(\lambda^*) = N \left[1 -
(1+\lambda^*) e^{-\lambda^*} \right] \quad ,\\
M_c(\gamma) &=& M - N t^* = N \left[ \frac1\gamma \left(
\frac{\lambda^*}{p} \right)^p \right]^\frac1{p-1} = \gamma N \left(1 -
e^{-\lambda^*} \right)^p  = N \frac{\lambda^*}{p} \left(1 -
e^{-\lambda^*} \right) \quad .
\end{eqnarray}
\end{widetext}
The first of these equations has a simple interpretation: The number
of vertices in the core is nothing but the number of vertices with a
degree larger than 1, after the application of the leaf removal
algorithm.  The second equation states that the number of hyperedges
left is the initial one minus the number of step the leaf removal
algorithm has been run (during each step only one hyperedge is
deleted).  The running time $t^*$ is the solution to
Eq.~(\ref{eq:lambda}) with $\lambda^*$ on the left hand side.  The
last two, and more compact, expressions for $M_c$ have been obtained
with the use of Eq.~(\ref{eq:ls}).  The lower curves in Fig.~\ref{p3}
show the normalized number of vertices $N_c/N$ and number of
interactions $M_c/N$ in the core as a function of $\gamma$, for $p=3$.

It is natural now to study the residual problem on the core,
$\hat{A}_c\,\vec{x}_c = \vec{y}_c \mod 2$, where $\hat{A}_c$ is the
$M_c \times N_c$ sparse random matrix obtained from $\hat{A}$ deleting
all the rows corresponding to removed interactions and all empty
columns.  In the next subsection we will derive a general result that,
when applied to the problem on the core, gives a necessary and
sufficient condition for the existence of solutions to
$\hat{A}_c\,\vec{x}_c = \vec{y}_c \mod 2$.  Then we will show that,
from a solution in the core, a solution for the original system can
always be constructed.
\vspace{1cm}

\subsection{Calculation of the $\gamma_c$ threshold}

Let us call $\mathcal{N}_{\mathcal{J},N,M}$ the number of GS for a
given disorder realization $\mathcal{J}$ (i.e.\ a given hypergraph and
coupling signs) with $N$ variables and $M$ interactions.  We will show
that, in the large $N$ limit, if the hypergraph does not contain any
vertex of degree less than 2, $\mathcal{N}_{\mathcal{J},N,M}$ is a
self averaging quantity, that is it does not fluctuate changing
$\mathcal{J}$.

In order to show self-averageness we will prove that, on hypergraphs
($p>2$) with minimum degree at least 2, the following equalities hold
\begin{equation}
\overline{\mathcal{N}_{\mathcal{J},N,M}} = 2^{N-M} \; , \;
\lim_{N \to \infty}
\frac{\overline{\mathcal{N}_{\mathcal{J},N,M}^{\;2}} -
\left(\;\overline{\mathcal{N}_{\mathcal{J},N,M}}\;\right)^2}
{\left(\;\overline{\mathcal{N}_{\mathcal{J},N,M}}\;\right)^2} = 0
\; , \label{eq:cond}
\end{equation}
where the overline stands for the average over the disorder ensemble,
that is over the ways of choosing $M$ hyperedges among $\binom{N}{p}$
and the ways of giving them a sign $J_m=\pm1$.  The above equalities
state that the probability distribution of
$\mathcal{N}_{\mathcal{J},N,M}$ over the disorder ensemble is a delta
function, and thus the quenched average equals the annealed one
\begin{equation}
\overline{\log \mathcal{N}_{\mathcal{J},N,M}} = \log
\overline{\mathcal{N}_{\mathcal{J},N,M}} = \log(2)\,(N-M) \quad .
\end{equation}
Given the definition
\begin{equation}
\mathcal{N}_{\mathcal{J},N,M} = \sum_{\vec\sigma} \prod_{m=1}^M
\delta(\sigma_{i_1^m} \ldots \sigma_{i_p^m} = J_m) \quad ,
\end{equation}
the first moment is trivially given by
\begin{equation}
\overline{\mathcal{N}_{\mathcal{J},N,M}} = \sum_{\vec\sigma}
\overline{\prod_{m=1}^M \delta(J_m = \sigma_{i_1^m} \ldots
\sigma_{i_p^m})} = 2^{N-M} \; ,
\end{equation}
since, for every given spin configuration and topology, the
probability that coupling signs satisfy all the $M$ interactions is
exactly $2^{-M}$.

The second moment is given by
\begin{widetext}
\begin{multline}
\overline{\mathcal{N}_{\mathcal{J},N,M}^{\,2}} =
\overline{\sum_{\vec\sigma\,\vec\sigma'} \prod_{m=1}^M
\delta(\sigma_{i_1^m} \ldots \sigma_{i_p^m} = J_m)
\delta(\sigma'_{i_1^m} \ldots \sigma'_{i_p^m} = J_m)} = \\
= \sum_{\vec\sigma} \overline{\prod_{m=1}^M \delta(J_m =
\sigma_{i_1^m} \ldots \sigma_{i_p^m}) \sum_{\vec\sigma'} \prod_{m=1}^M
\delta(\sigma'_{i_1^m} \ldots \sigma'_{i_p^m} = \sigma_{i_1^m} \ldots
\sigma_{i_p^m})} = 2^{N-M} \overline{\sum_{\vec\tau} \prod_{m=1}^M
\delta(\tau_{i_1^m} \ldots \tau_{i_p^m} = 1)} \quad ,
\label{eq:mom2}
\end{multline}
where $\tau_i = \sigma_i \sigma'_i$ and the last expression is nothing
but the annealed average of the partition function at $T=0$ for a
system where all the coupling signs have been set to 1, i.e.\ a
ferromagnetic model.  Such an average can be computed by standard
saddle point integration and the final result is
\begin{equation}
F_{N,M} = \lim_{N \to \infty} \frac1N \log \overline{\sum_{\vec\tau}
\prod_{m=1}^M \delta(\tau_{i_1^m} \ldots \tau_{i_p^m} = 1)} = \sum_k
P(k)\; \log\left(x_+^k + x_-^k\right) \quad ,
\label{eq:entro}
\end{equation}
\end{widetext}
where $P(k)$ is the distribution of connectivities in the hypergraph
and $x_+,x_-$ solve the following equations
\begin{eqnarray}
x_+ + x_- = \left[ \sum_k \frac{k P(k)}{\langle k \rangle} \;
\frac{x_+^{k-1} + x_-^{k-1}}{x_+^k + x_-^k} \right]^{p-1} \quad ,
\label{eq:spe1} \\
x_+ - x_- = \left[ \sum_k \frac{k P(k)}{\langle k \rangle} \;
\frac{x_+^{k-1} - x_-^{k-1}}{x_+^k + x_-^k} \right]^{p-1} \quad .
\label{eq:spe2}
\end{eqnarray}
Here $\langle k \rangle = \sum_k k P(k) = p \frac{M}{N}$ is the mean
connectivity.  When more than one solution to
Eqs.(\ref{eq:spe1},\ref{eq:spe2}) exist, the one maximizing
Eq.~(\ref{eq:entro}) must be chosen.  The value of $x_+$ (resp. $x_-$)
is proportional to the fraction of variables taking values 1
(resp. -1) in the set of configurations which maximize the sum in
Eq.~(\ref{eq:entro}).  Then the typical magnetization of this model is
given by $m = \frac{x_+ - x_-}{x_+ + x_-}$.

Solutions to Eqs.(\ref{eq:spe1},\ref{eq:spe2}) can be classified
depending on the value of magnetization $m$.  In full generality there
are 3 solutions: a first symmetric one ($x_+=x_-$) with $m=0$, a
second one with large magnetization and a third one with an
intermediate value of $m$.  For some choices of $P(k)$ (e.g.\ a
Poissonian) solutions with $m>0$ may exist only for $\frac{M}{N}$
large enough.  The solution with intermediate magnetization always
corresponds to a minimum of $F_{N,M}$ and can be in general neglected.

The symmetric solution $x_+=x_-=2^{-1/p}$ always exists and gives
$F_{N,M} = \log(2)\,(1-\frac{M}{N})$.  For $p>2$ and $P(0)=P(1)=0$,
i.e.\ for hypergraphs with minimum degree 2, the solution with large
magnetization also exist for any $\gamma$ value and has $x_+=1$,
$x_-=0$ and $F_{N,M}=0$.  As expected, the intermediate solution, when
it exists, has $F_{N,M}<0$.

Then, for $p>2$ and $P(0)=P(1)=0$, we can conclude that the average in
the last term of Eq.~(\ref{eq:mom2}) equals $e^{N F_{N,M}} = 2^{N-M}$
(the coefficient can be easily calculated and is exactly 1).  Thus,
equalities in Eq.~(\ref{eq:cond}) hold and the number of GS is a
self-averaging quantity.

Since the core generated by the leaf removal algorithm has minimum
degree 2, we may apply the above result, and find that the SAT/UNSAT
threshold is given by the condition
\begin{multline}
N_c(\gamma_c) = M_c(\gamma_c) \quad \Longrightarrow \\
\Longrightarrow \quad 1 - (1 + \lambda_c) e^{-\lambda_c} =
\frac{\lambda_c}{p} \left( 1 - e^{-\lambda_c} \right) \quad ,
\label{eq:gc}
\end{multline}
where $\lambda_c = \lambda^*(\gamma_c)$.  For $\gamma \le \gamma_c$
there are $2^{N_c-M_c}$ solutions (i.e.\ unfrustrated GS) in the core,
while for $\gamma > \gamma_c$ there is none.  For $p=3$, solution to
Eq.~(\ref{eq:gc}) gives $\lambda_c = 2.14913$ and $\gamma_c =
0.917935$.

For any given solution in the core, a solution for the whole original
system can be easily reconstructed.  Indeed, we reintroduce in the
system the interactions removed during the leaf removal process, but
in a reversed order (i.e.\ the last removed is the first to be
reintroduced).  At each step, together with one interaction, at least
one variable is reintroduced in the system (the variable having degree
1 when that interaction was removed) and this variable must be set
such as to satisfy the interaction.  Very often more than one variable
per step is reintroduced, allowing for multiple and equivalent
choices.  This redundancy is what makes the total number of solutions
larger than the number of solutions in the core (see below).

In the table below we report the thresholds $\gamma_d$ and $\gamma_c$
for some $p$ values.

\begin{center}
\begin{tabular}{|c|c|c|}
\hline
\ $p$ \ & $\gamma_d$ & $\gamma_c$ \\
\hline \hline
2 & 1/2 & 1/2 \\
\hline
3 & \ 0.818469 \ & \ 0.917935 \ \\
\hline
4 & \ 0.772278 \ & \ 0.976770 \ \\
\hline
5 & \ 0.701780 \ & \ 0.992438 \ \\
\hline
6 & \ 0.637080 \ & \ 0.997380 \ \\
\hline
\end{tabular}
\end{center}

\subsection{Clustering of ground states}

Let us come back to the problem of clustering solutions before the
SAT/UNSAT threshold ($\gamma \le \gamma_c$).  In this region the
system is not frustrated and then a gauge transformation setting all
coupling signs to 1 can always be found: Given an unfrustrated GS
$\vec\sigma^0$ a possible gauge transformation is $\sigma'_i =
\sigma_i \sigma^0_i$ and $J'_m = J_m \sigma^0_{i_1^m} \ldots
\sigma^0_{i_p^m} = 1$.  Thanks to this, in the rest of the paper we
will consider only a ferromagnetic system ($J_m=1 \; \forall m$),
which corresponds to the linear system $\hat{A}\,\vec{x} = \vec{0}
\mod 2$.

The solutions to the linear system $\hat{A}\,\vec{x} = \vec{0} \mod 2$
form a group: The sum of 2 solutions is still a solution and the null
element is the solution $\vec{x} = \vec{0}$.  The symmetry group is
telling us that if one looks at the configurational space sitting on a
reference GS, the set of GS will look the same, whatever the reference
GS is.  An immediate consequence of this symmetry is that, if GS form
clusters, these clusters must be all of the same size.

For $\gamma \le \gamma_c$, hyperloops are absent and the total number
of GS (or solutions) is always given by $2^{N-M}$, i.e.\ their entropy
is $S(\gamma) = \log(2)\;(1-\gamma)$.  Let us divide the $N$ variables
in 2 sets: $\vec{x}_c$ represents the $N_c$ variables in the core, and
$\vec{x}_{nc}$ the $N-N_c$ variables in the non-core part of the
hypergraph, that is variables corresponding to vertices remained
isolated at the end of the leaf removal process.  Thus also the
entropy can be divided in 2 parts.  One part is given by the solutions
in the core, that is by the possible assignments of $\vec{x}_c$,
\begin{equation}
S_c(\gamma) = \log(2) \frac{N_c(\gamma)-M_c(\gamma)}{N} \quad ,
\end{equation}
which is non-negative for $\gamma_d \le \gamma \le \gamma_c$.  The
other part is given by the possible multiple assignments of
$\vec{x}_{nc}$ during the reconstruction process
\begin{equation}
S_{nc}(\gamma) = S(\gamma) - S_c(\gamma) \quad .
\end{equation}
This separation of the entropy in 2 parts is physically relevant, and
we will show here that it corresponds to the proper clustering of the
solutions.

\begin{figure}[!ht]
\includegraphics[width=\columnwidth]{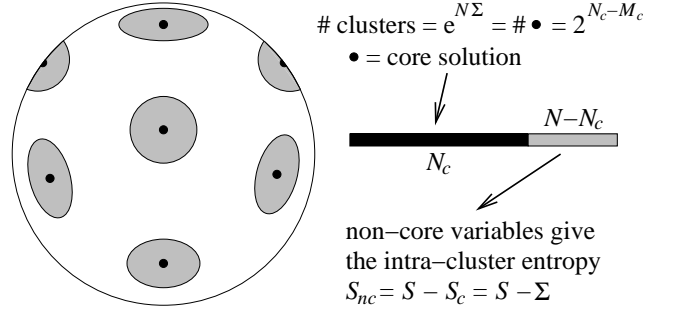}
\caption{Schematic picture of the clustering of solutions for
$\gamma_d < \gamma < \gamma_c$.}
\label{clust}
\end{figure}

The physical picture we have in mind is sketched in Fig.~\ref{clust}.
For $\gamma_d \le \gamma \le \gamma_c$, the solutions of
$\hat{A}\,\vec{x} = \vec{y} \mod 2$, or equivalently the ground states
of (\ref{eq:ham}), spontaneously form clusters.  By definition, two
solutions having a finite Hamming distance $d$, i.e.\ $d/N \to 0$ for
$N \to \infty$, are in the same cluster, while two solutions in
different clusters must have an extensive distance, that is $d/N \sim
\mathcal{O}(1)$ for large $N$.

\begin{figure}[!ht]
\includegraphics[width=\columnwidth]{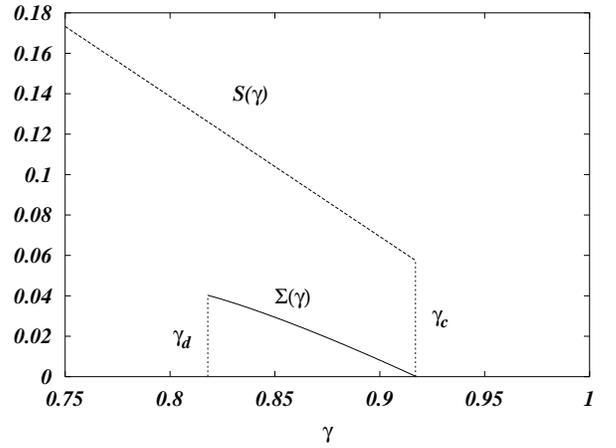}
\caption{Total entropy $S(\gamma)$ and configurational entropy
$\Sigma(\gamma)$ for $p=3$.}
\label{entropie}
\end{figure}

In virtue of the property stated at the beginning of this subsection,
all the clusters have the same size.  Their number is $e^{N
\Sigma(\gamma)}$, where $\Sigma(\gamma)$ is called complexity or
configurational entropy.  We will show that the number of clusters
equals the number of solution in the core, that is
\begin{equation}
\Sigma(\gamma) = S_c(\gamma) \quad .
\label{eq:complex}
\end{equation}
The intra-cluster entropy, i.e.\ the normalized logarithm of the
cluster size, is then given by the non-core entropy $S_{nc}(\gamma) =
S(\gamma) - S_c(\gamma) = S(\gamma) - \Sigma(\gamma)$.  For $p=3$
these entropies are shown in Fig.~\ref{entropie}.

The proof of Eq.~(\ref{eq:complex}) is given in 2 steps.  First we
show that all the solution assignments of the core variables
$\vec{x}_c$ are ``well separated'', that is the distance among any
pair of them is extensive.  This is what gives rise to the clustering,
with a number of clusters which is at least as large as the number of
core solutions ($\Sigma \ge S_c$).  Then we show that, for any fixed
$\vec{x}_c$, all possible assignments of non-core variables
$\vec{x}_{nc}$ belong to the same cluster, and so $\Sigma = S_c$.

The first step is accomplished by calculating the probability
distribution of the distance among any two solutions in the core.
Thanks to the group property, we can restrict the calculation fixing
one solution to the null vector $\vec{0}$.  For simplicity we have
performed an annealed average
\begin{widetext}
\begin{equation}
S(d,\gamma) = \lim_{N_c \to \infty} \frac{1}{N_c} \log
\overline{\sum_{\vec\sigma} \delta\Bigl(\sum_i \sigma_i = N_c - 2
d\Bigr) \prod_{m=1}^{M_c} \delta(\sigma_{i_1^m} \ldots \sigma_{i_p^m}
= 1)} \quad ,
\end{equation}
\end{widetext}
which gives an upper bound to the exact result.  The expression for
this entropy is given by Eq.~(\ref{eq:entro}), where $x_+$ comes from
the solution of Eq.~(\ref{eq:spe1}), keeping the ratio
$\frac{d}{N_c}=\frac{x_-}{x_+ + x_-}$ fixed.

\begin{figure}[!ht]
\includegraphics[width=\columnwidth]{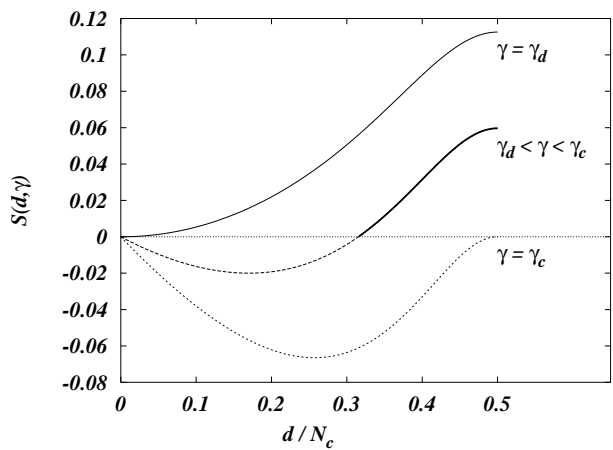}
\caption{Entropy of distances among solutions in the core for $p=3$
(in the annealed approximation).}
\label{entro_ann}
\end{figure}

\begin{figure}[!ht]
\includegraphics[width=\columnwidth]{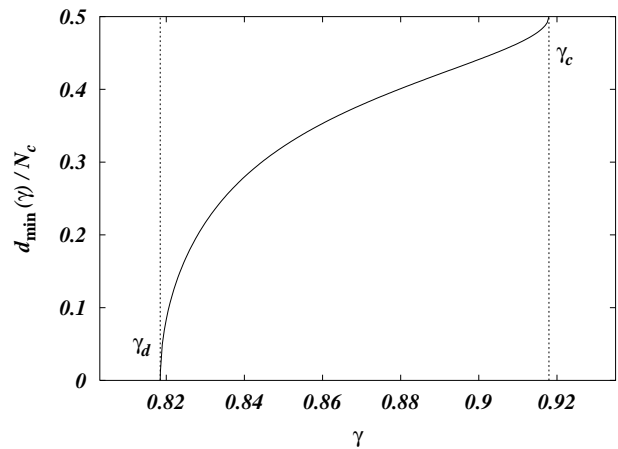}
\caption{Lower bound for the minimum distance among any 2 solutions in
the core for $p=3$.}
\label{min_dist}
\end{figure}

In Fig.~\ref{entro_ann} we plot the resulting entropy as a function of
the distance $d$, for $p=3$ and some values of $\gamma$.  For
$\gamma_d < \gamma < \gamma_c$ the entropy is negative for $0 < d <
d_{\text{min}}(\gamma)$, and so $d_{\text{min}}(\gamma)$ is a lower
bound on the minimum distance among any two solutions in the core.
This minimum distance is shown for $p=3$ in Fig.~\ref{min_dist}.

Then all the $e^{N S_c}$ core solutions are well separated, and can be
represented as the centers of the clusters (see Fig.~\ref{clust}).  It
remains to be proven that, for any fixed $\vec{x}_c$, the solution
assignments of $\vec{x}_{nc}$ form a single cluster.  Thus no further
clustering is present and the picture of Fig.~\ref{clust} is correct.

This last proof is given in the Appendix, and it is based on an
algorithm which allows one to change the value to any variable in
$\vec{x}_{nc}$ by simply adjusting other $\mathcal{O}(1)$ variables in
$\vec{x}_{nc}$.  This shows that all the solutions in one cluster are
connected in the following sense. One solution can be reached from any
other one by a sequence of moves, where each move involves flipping
only a finite number of spins.


\section{Conclusion and discussion}

In this work we have solved, with two alternative methods, the
$p$-XORSAT model, which corresponds to the zero-temperature limit of
the diluted $p$-spin model.

Increasing the $\gamma$ parameter (number of interactions per
variable) the model undergoes two phase transitions.  At $\gamma_d$,
solutions to the $p$-XORSAT problem (i.e. ground states for the
$p$-spin model) spontaneously form an exponentially large number of
clusters, thus giving a finite configurational entropy.  At
$\gamma_c$, frustration percolates throughout the system, and
consequently the number of clusters (and solutions) goes to zero, and
the ground state energy becomes positive.  $\gamma_c$ corresponds to
the SAT/UNSAT threshold.  These exact results perfectly agree with
previous replica calculations \cite{RIWEZE,LERIZE,FLRZ} and may
suggest new approaches for finding mathematical bases to Parisi's
theory of spin glasses \cite{MEPAVI}.

The use of the cavity method combined with a rigorous derivation based
on the topological properties of the interaction hypergraph, allow us
to establish some interesting links among distributions of cavity
fields on a given variable and the position of the corresponding
vertex in the hypergraph.  In particular all the variables with a
non-trivial distribution of cavity fields belong to the `frozen' part
of the hypergraph (see Appendix), that is to the core and to the part
that can be uniquely fixed, once an assignment to core variables has
been chosen.  The `frozen' part is exactly the backbone of a cluster
(variables which take the same value for all the solutions in the
cluster) and its size is given by the largest solution to
Eq.~(\ref{eq:mag}).  The rest of the hypergraph, the `floppy' part,
only contains paramagnetic variables, that is variables always having
a null cavity field.


\begin{acknowledgments}

We thank Silvio Franz, Michele Leone and Martin Weigt for discussions.
F.R.T.\ thanks ICTP for kind hospitality during the completion of this
work.

While writing this work, we became aware of an independent work by
S. Cocco, O. Dubois, J. Mandler and R. Monasson on similar issues
\cite{CDMM}.

\end{acknowledgments}


\appendix
\section{}

In this appendix we show that assignments of non-core variables
$\vec{x}_{nc}$ are not clustered.  To this end, we define an algorithm
which allows one to flip any non-core variable, by simply adjusting
other $\mathcal{O}(1)$ non-core variables.  With this algorithm one
can move through all the $\vec{x}_{nc}$ assignments by doing {\em
finite} steps, thus proving that non-core solutions form a single
cluster.

\begin{figure}[!ht]
\includegraphics[width=0.9\columnwidth]{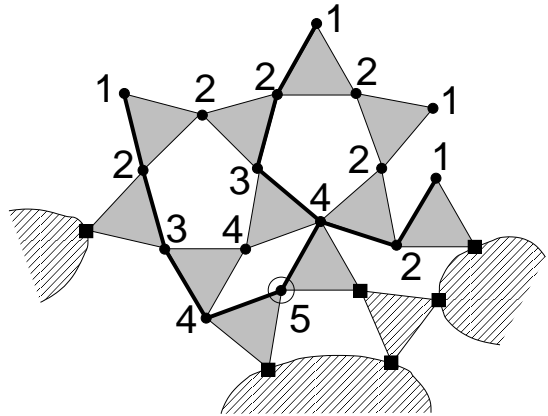}
\caption{The bold tree-like structure is a possible seaweed (see text)
in order to flip the variable on the circled vertex and still keep all
the interactions satisfied. Note that the seaweed passes through at
most 2 vertices on the same interaction.}
\label{nc}
\end{figure}

Let us fix the core variables $\vec{x}_c$ to any solution, and call
them `frozen'.  All the variables, belonging to at least one equation
where the other $p-1$ variables are already frozen, must be frozen too
(see e.g.\ the dashed triangle in Fig.~\ref{nc}, where the dashed
blobs represent the frozen core).  In this way one is able to freeze a
number of variables $m(\gamma) N$, where $m(\gamma)$ turns out to
coincide with the largest solution of Eq.~(\ref{eq:mag}), that is with
the magnetization in the ferromagnetic state or the backbone in a
generic cluster.  For $p=3$ the function $m(\gamma)$ is shown in
Fig.~\ref{p3} (upper curve).

After having fixed all the variables one could, one is left with the
`floppy' part of the hypergraph.  The typical situation is sketched in
Fig.~\ref{nc}, where the dashed part is frozen (hereafter we refer
only to the $p=3$ case for the sake of clarity).  All the interactions
involving both frozen and floppy variables (those which form the
boundary between the frozen and the floppy part of the hypergraph)
must contain 2 floppy and 1 frozen variables, otherwise (2 frozen and
1 floppy) that interaction would become frozen as well and would not
longer be on the boundary.

The numbers in Fig.~\ref{nc} have been assigned during a slightly
different leaf removal process with the following rule.  Starting with
the original hypergraph, the number ``1'' is given to all the vertices
of degree less than 2 (isolated vertices and leafs) and their
hyperedges are deleted.  Then, in the new hypergraph, the number ``2''
is given to all vertices of degree less than 2 and their hyperedges
deleted.  And so on.  We call these numbers the {\em depth} of a
vertex: Vertices of depth 1 represent the `external boundary' or the
`surface' of the hypergraph.

The evolution of this ``collective'' leaf removal process can be
described in terms of the same function $f_1(t)$ used previously.  At
each time step a depth is assigned to a fraction $f_1(t)$ of vertices
and then the time is increased by $\Delta t = f_1(t)$, in order to
take into account the deletion of hyperedges leaving from the just
numbered vertices.  For very large times and depths, $f_1(t)$ is very
small and can be approximated by $f_1(t) \simeq (t-t^*) \partial_t
f_1(t^*)$, where $t^*$ is such that $f_1(t^*)=0$.  In this regime we
have that
\begin{eqnarray}
\Delta t &=& f_1(t) \quad , \\
\Delta f_1(t) &=& \left. \frac{\partial f_1(t)}{\partial t}
\right|_{t^*} \Delta t = \left. \frac{\partial f_1(t)}{\partial t}
\right|_{t^*} f_1(t) \quad ,
\end{eqnarray}
and so $f_1(t+\Delta t) = f_1(t) \left[1 + \partial_t f_1(t^*)
\right]$.  Then the probability of having a (large) depth $h$ satisfy
the equation $\mathcal{P}(h+1) \simeq \mathcal{P}(h)\; \mu$, where
\begin{equation}
\mu(\gamma) = 1 + \left. \frac{\partial f_1(t)}{\partial t}
\right|_{t^*} = 1 + \left. \frac{\partial f_1(\lambda)} {\partial
\lambda} \frac{\partial \lambda(t)}{\partial t}
\right|_{\lambda^*(\gamma)} \; .
\end{equation}
Since the probability of having depth $h$ drops exponentially for
large $h$ as $\mathcal{P}(h) \propto \mu^h$, the largest depth
assigned with this process is $\mathcal{O}(\log N)$.  For any $\gamma
\neq \gamma_d$ we have that $\mu(\gamma) < 1$, since $\lambda(t)$ is a
decreasing function of $t$ and $\partial_\lambda f_1(\lambda)$ is
positive in the largest root $\lambda^*$, unless $\gamma=\gamma_d$.

Once depths have been assigned, there is an algorithm (described
below) which allows one to change the value to any floppy variable, by
adjusting, at the same time, only $\mathcal{O}(1)$ other floppy
variables.  Such a new configuration will be a finite distance far
apart, and, by definition, will belong to the same cluster.  In this
way one can change the configuration of the floppy (and non-core)
variables to any admissible one, and these configurations will form a
unique cluster.

The physical idea behind the algorithm for flipping any floppy
variable, keeping all the interactions satisfied, and adjusting only a
finite number of other variables, is the following.  Suppose, as in
Fig.~\ref{nc}, that we flip the variable of depth 5.  Then the
interactions it participates to will be unsatisfied, and we have to
move this `excess energy' by flipping other variables, along the
shortest way, towards the boundaries of the hypergraph, that is the
vertices of depth 1, where it can be freely released.  The only
delicate point is the definition of the `path to the boundary', which
has to contain a finite number of vertices.  In Fig.~\ref{nc} we show
a possible way to release the excess energy generated by flipping
variable of depth 5: Flipping all the variables belonging the
tree-like bold structure will keep all the interactions satisfied,
since every interaction contains an even number of vertices belonging
to the bold structure.

We will call this tree-like structure a {\em seaweed}, since it has a
root, corresponding to the vertex of maximum depth, and the number of
its branches grows approaching the surface.  Now we give the rules for
constructing a seaweed, such that its size is finite.

Let us start with some nomenclature: We say that a hyperedge $e$ is
``below'' a vertex $v$, and analogously $v$ is ``above'' $e$, if the
depth of $v$ is the smallest among the depths of all the vertices in
$e$.

Thanks to the way depths have been assigned, each vertex may have at
most 1 hyperedge below.  This property can be easily proved,
remembering that to any given vertex $v$ the depth is assigned only
when its connectivity is 0 or 1.  At this time, all the other
hyperedges of $v$ have been removed, since we have assigned smaller
depths to its neighbours.  The only hyperedge which can be below $v$
is the last one.  Moreover, if the depth is assigned to $v$ when its
connectivity is 0 (isolated vertex), the vertex $v$ will have no
hyperedges below, and we will call it a {\em root}.

The construction of the seaweed starts from the vertex corresponding
to the variable that we want to flip (let us call it {\em seed}).  In
this way we are sure that such a vertex will be in the structure, and
the corresponding variable flipped.  The seaweed is built up
recursively, that is we give the rules for growing a single branch,
both upwards (i.e.\ towards the surface) and downwards (i.e.\ towards
a root), and then these rules must be applied to any branch of the
seaweed, until it reaches the surface of the hypergraph or a root
vertex.  The branches are such that along an upwards (downwards)
direction the depth strictly decreases (increases).  Rare exceptions
to this property will be illustrated below.

When a branch passes through a hyperedge it will visit only 2 vertices
in this hyperedge, such that, when all the variables belonging to the
seaweed will be flipped, the interaction will remain satisfied.

Suppose the seed vertex has connectivity $k$.  Then we start $k$
different branches, 1 downwards entering the only hyperedge below the
seed vertex and $k-1$ upwards entering the other hyperedges.

Any upwards branch entering a hyperedge $e$ through vertex $v$ has to
be continued with the vertex above $e$.  If there are many vertices of
the same minimum depth in $e$, any of them can be chosen equivalently.
With this rule we are ensuring that the new vertex added to the
upwards branch is of smaller depth than $v$.

Any downwards branch entering hyperedge $e$ through vertex $v$ has to
be continued with the vertex of maximum depth in $e$.  If there are
many vertices of the same maximum depth in $e$, any of them can be
chosen equivalently.  With this rule we can ensure that the new vertex
added to the downwards branch will be deeper than $v$, since $v$ is of
minimum depth in $e$.

Any growing branch reaching a vertex $v$ of connectivity $k$ has to be
continued with $k-1$ branches, in order to satisfy the rule that all
the hyperedges of $v$ must be visited by a branch.  If the just
reached vertex is on the surface (i.e.\ it has depth 1 and
connectivity 1) the branch ends there.  On the contrary, reaching a
vertex of connectivity larger than 2, the growing branch generates new
branches.  More in particular, if the branch is an upwards one it will
generate only upwards branches (since it is coming from the only
hyperedge below $v$).  While, if it is the downwards one, it may
generate at most one downwards branch (all the rest being upwards
ones).  This is a consequence of the property that every vertex may
have at most 1 hyperedge below it.

In two cases the unique downwards branch ends in a vertex $v$, which
is thus the root of the seaweed: (1) $v$ is a root vertex, that is it
has no hyperedges below it (2) vertex $v$ is above hyperedge $e$, but
$v$ is not the only vertex of minimum depth in $e$.  In this case the
branch entering $e$ through $v$ becomes an upwards one, and makes a
single step without decreasing the depth (this is the only exception
to the rule on the monotonicity of the depth along a branch stated
above).

Since each branch of the seaweed is grown independently, it may be
that a the end of the process some vertices result in more than one
branch.  This is not a problem: The rule says that every vertex which
has been included an odd number of times in the seaweed must be in it;
While those entering an even number of times must be left out.  The
net result is a decrease in the total number of vertices in the
structure.  The seaweed can eventually break up in more than a single
connected component: All the components, but that containing the seed,
can be removed from the seaweed.

The choice of growing the branches always along vertices of maximum
and minimum depths is dictated by the need of reaching a root vertex
and the surface of the hypergraph as soon as possible, thus making the
seaweed as small as possible.  It is worth noticing that the
probability that a vertex is a root increases for larger depths.

The last point to be proven is that the typical distance, $\ell$,
measured along any branch, among the root of the seaweed and the
surface, is finite (and not order $\log N$).  This property together
with the fact that the branching ratio is proportional to the
connectivity, which is finite too, implies that the number of vertices
in the seaweed, which is roughly proportional to $(3\gamma-1)^\ell$,
is finite.  On the contrary if $\ell$ would be of order $\log N$, the
volume of the seaweed would diverge for large $N$.

In order to show that $\ell$ is finite, even when the root depth is as
large as possible (i.e.\ order $\log N$), we need to know the
probability that a vertex has depth $h$.  This probability
distribution function, $\mathcal{P}(h)$, can be calculated exactly,
but its expression is too involved to be presented here.  We only
report some features relevant for our purposes.  It depends on the
connectivity of the vertex, $\mathcal{P}_k(h)$, and for $k=0$ or $k=1$
it is trivially given by $\mathcal{P}_{0,1}(h)=\delta(h-1)$.  For any
$k \ge 2$, it decreases exponentially fast for large $h$, and the
probability of reaching a vertex (not on the surface) of depth $h$ is
$Q(h) = \sum_{k \ge 2} k f_k(0) \mathcal{P}_k(h) \propto
\mu(\gamma)^h$ for large $h$.  For the present calculation the exact
shape of $Q(h)$ at small depths is irrelevant, and we only care about
its tail, so we can hereafter use $Q(h) = \mu^h$ for all $h$.

We show now that, with such a distribution of depths, even starting
from a root of depth $\mathcal{O}(\log N)$, an upwards branch needs
only a finite number of steps to reach the surface (for simplicity we
fix to 0, instead of 1, the surface depth).  The probability of going
in a single step from depth $h_1$ to depth $h_2$ is
\begin{equation}
w(h_1 \to h_2) = \frac{1-\mu\ }{1-\mu^{h_1}} \mu^{h_2} \quad ,
\end{equation}
which has the correct normalization $\sum_{h_2=0}^{h_1-1} w(h_1 \to
h_2)=1$.  The probability of going from depth $h$ to depth 0 in $m$
steps is then
\begin{widetext}
\begin{multline}
W_h(m) = \sum_{h_1=h_2+1}^{h-1} \sum_{h_2=h_3+1}^{h_1-1} \ldots
\sum_{h_{m-1}=1}^{h_{m-2}-1} w(h \to h_1) w(h_1 \to h_2) \ldots
w(h_{m-2} \to h_{m-1}) w(h_{m-1} \to 0) = \\
= \frac{1-\mu}{1-\mu^h} \frac{(1-\mu)^{m-1}}{(m-1)!} \sum_{\{i\}}
\mbox{}^\prime \prod_{j=1}^{m-1} \frac{\mu^{i_j}}{1-\mu^{i_j}} <
\frac{1-\mu}{1-\mu^h} \frac{(1-\mu)^{m-1}}{(m-1)!} \left[
\sum_{i=1}^{h-1} \frac{\mu^i}{1-\mu^i} \right]^{m-1} \quad ,
\end{multline}
\end{widetext}
where the primed sum is over the $m-1$ intermediate depths, taking
different values between 1 and $h-1$, and the inequality follows since
in the last term we have included also configurations with indices
taking equal values.  So $W_h(m)$ is upper bounded by a Poissonian
distribution with a mean number of steps
\begin{equation}
\ell(h) = (1-\mu) \sum_{i=1}^{h-1} \frac{\mu^i}{1-\mu^i} \quad .
\end{equation}
As expected, $\ell$ is an increasing function of $h$.  In the limit of
a very deep root, $h \to \infty$, the series converges for any $\mu<1$
(i.e.\ $\gamma>\gamma_d$), and thus $\ell(\infty)$ is still finite.


\end{document}